\documentclass[a4paper]{aa}
\usepackage{graphicx}
\begin{document}
   \title{Seed fields for galactic dynamos by the magnetorotational instability}

   \titlerunning{MRI in galaxies}
\authorrunning{L.L. Kitchatinov \& G. R\"udiger}
   \author{L. L. Kitchatinov\inst{1,2} \and G. R\"udiger\inst{1}}


   \institute{Astrophysikalisches Institut Potsdam, An der Sternwarte 16,
              D-14482, Potsdam, Germany \\
              \email{gruediger@aip.de; lkitchatinov@aip.de}
         \and
             Institute for Solar-Terrestrial Physics, PO Box
             4026, Irkutsk 664033, Russian Federation\\
             \email{kit@iszf.irk.ru}
             }

   \date{\today }

\abstract{A linear but global numerical model for the magnetorotational
instability (MRI) in disk geometry is considered to estimate the instability
parameters for galaxies. Similarity rules suggested by a local analysis
are applied to reveal a universal behavior approached by the
results of global calculations for large magnetic Prandtl
numbers. The findings are used to estimate the MRI
characteristics for galaxies with their very large magnetic
Prandtl numbers which cannot be attained in any numerical simulations.
The resulting minimum field for the instability, $B_{\rm min} \simeq
10^{-25}$~G, is small compared to any seed fields currently
discussed. The growth times  of MRI are estimated to be on the order of
the rotation period of the inner rigidly-rotating core,
i.e. $\sim$100 Myr. Global MRI excites preferentially the magnetic
field modes of quadrupolar symmetry.
}
\maketitle
\section{Introduction}
The magnetorotational instability (MRI) has been discovered
by Velikhov (1959) for hydromagnetic Couette flows;  in the astrophysical
context it has been established by Fricke (1969) and  Balbus \& Hawley
(1991). Several properties of the instability make it also  relevant for
the theory of galactic magnetism.

The instability  exists due to
magnetic fields, but the minimum field for MRI is really very small. We
shall see that for galactic conditions the minimum field, $B_{\rm
min} \simeq 10^{-25}$~G, is much smaller than the seed
magnetic fields of protogalaxies whatever origin of the seed
fields we assume. With its growth time of order the rotation time the
MRI is fast, with e-folding times below
$0.1$~Gyr. The observation of Faraday rotation of high-redshift objects by
Athreya et al. (1998) and Carilli \& Taylor (2002) seems to indicate the real
existence of such short growth times of  fields (Lesch \& Hanasz 2003).

It was recognized by Beck et al. (1994) that the seed field can be
largely amplified by interstellar turbulence before the global
hydromagnetic dynamo starts operating (Cattaneo 1999, Schekochihin et al. 2004).
Turbulent velocities of about 10 km/s are indeed
indicated by spectral line broadening of protogalactic disks
(Wolfe 1988). The origin of the turbulence is, however, uncertain.
Supernova explosions would drive a much weaker turbulence
(Ferri\`{e}re 1992, Kaisig et al. 1993). Another  driver is MRI which was
demonstrated by Sellwood \& Balbus (1999) to produce its own turbulence
in the nonlinear regime.

A more specific problem of galactic dynamos to which MRI can also
be relevant is the parity selection by the global magnetic dynamo.
Galactic dynamo models prefer the global fields symmetric about the disk
midplane (Stix 1975, and many other references in R\"udiger \& Hollerbach 2004)
which symmetry type is favored by observations also. In order to be amplified,
however, the quadrupolar field should already be present in the seed field for the dynamo. Not all origins for the seed field currently discussed lead to the
required symmetry, e.g., battery effect of Biermann supplies dipolar global
fields (Krause \& Beck 1998). We shall demonstrate that MRI excites
preferentially global modes with quadrupolar parity even if the (axial)
destabilizing field is antisymmetric with respect to the disk midplane
which property suggests MRI as a natural supplier of the seed field
for global galactic dynamos.

One could argue that MRI in the weak-field regime is rather local
by nature and not efficient in producing global patterns.
The local theory of ideal fluids yields the growth rate, $\gamma$,
for MRI proportional to the Alfv\'en frequency, $\gamma \sim
V_{\rm A}k$, in the weak-field case, $V_{\rm A} \ll \Omega/k$
($V_{\rm A}=B_0/\sqrt{\mu_0 \rho}$ is the Alfv\'en
velocity, $k$ is the vertical wave number, $\Omega$ is the local
angular velocity, and $B_0$ is the background axial magnetic field).
The growth rate seemingly increases with the wave number.
The growth is, however, stabilized by the microscopic magnetic
diffusivity, $\eta$, so that $\gamma + \eta k^2 \simeq V_{\rm
A}k$. Marginal stability then demands $V_{\rm A} \simeq \eta k$.
The smallest field strength producing the instability belongs to
the smallest possible $k$. The low bound on magnetic field
producing MRI is therefore controlled by the global geometry.

We  try to find
out in the present paper whether MRI should be expected for protogalaxies
where only very weak seed fields are present and the magnetic Prandtl number is
extremely large (see Kulsrud \& Anderson 1992). The extent to which the
seed field can be amplified by the turbulence is
evaluated. The preferred geometry of magnetic field in the global
unstable modes is determined. To this end the disk model of
Kitchatinov \& Mazur (1997) is used which is outlined in the next Section.
Section~\ref{results} presents the results of the computations.
The implications for galactic dynamos are discussed in the final
Section~\ref{discussion}.
\section{The model}
The model concerns a rotating disk of constant thickness, $2H$,
threaded by a uniform axial magnetic field. The rotation axis is
normal to the disk and the angular velocity, $\Omega$, depends on
the distance, $s$, to the axis. This dependence is parameterized
by
\begin{equation}
   \Omega (s) = \Omega_0 \tilde\Omega (s)
\label{1a}
\end{equation}
with
\begin{equation}
   \tilde\Omega (s) = \left( 1 + \left( {s\over s_0}\right)^n\
   \right)^{-1/n} .
\label{1}
\end{equation}
This profile describes almost uniform rotation up to the distance $s_0$,
and the law $\Omega \simeq \Omega_0 (s_0/s)$, implying constant rotation
velocity for distances
$s \gg s_0$, as is  observed for nearly all  galaxies.
We use $n=2$ in Eq.~(\ref{1}) and
 $s_0/H = 5$ for the aspect ratio.

We assume the fluid be incompressible, ${\rm div}\ {\vec U} = 0$. The fluid compressibility is probably not important for our analysis because MRI proceeds via noncompressive Alfv\'en-type disturbances. The magnetic field and the flow in the disk obey the induction equation,
\begin{equation}
   {\partial{\vec B}\over\partial t} = {\rm curl}
   \left({\vec U}\times{\vec B}\right) + \eta\Delta{\vec B} ,
\label{2}
\end{equation}
and the equation for vorticity, ${\vec{\cal W}} = {\rm curl}\,{\vec U}$,
\begin{equation}
   {\partial{\vec{\cal W}}\over\partial t}  = {\rm curl}\left({\vec U}\times
   {\vec{\cal W}} + {\vec J}\times{\vec B}/\rho \right)
   + \nu\Delta{\vec{\cal W}} ,
\label{3}
\end{equation}
which results from taking the curl of the momentum equation to eliminate the pressure.
${\vec J} = (1/\mu_0)\ {\rm curl}\,{\vec B}$ is the current
density.

To perform the linear stability analysis, we linearize (\ref{2})
and (\ref{3}) about the rotation (\ref{1}) and the
uniform axial field, ${\vec B}_0 = B_0 \hat{\vec{e}}_z$ ($\hat{\vec{e}}_z$ is
the unit vector along the rotation axis), and introduce the
normalized variables,
\begin{eqnarray}
\lefteqn{   {\vec b} = {\vec B}/B_0,\ \, {\vec j} = {\mu_0 H\over  B_0}{\vec
J},\ \,
   {\vec u} = {\vec U}/(H\Omega_0), \ \,
      {\vec\omega} = {\vec{\cal W}}/\Omega_0} .
 \label{4}
\end{eqnarray}
This leads to the three basic dimensionless parameters of magnetic Reynolds
number, Rm, Hartmann number, Ha, and  magnetic Prandtl number, Pm,
\begin{equation}
   {\rm Rm} = {\Omega_0 H^2\over\eta},
   \ \ \ \
   {\rm Ha} = {B_0 H\over \sqrt{\mu_0\rho\nu\eta}} = {V_{\rm A} H\over\sqrt{\nu\eta}},
   \ \ \ \
   {\rm Pm} = {\nu\over\eta} .
   \label{7}
\end{equation}
They  control the equation system for the normalized disturbances,
\begin{eqnarray}
   {\partial{\vec b}\over\partial t} &=& {\rm Rm\ {\rm curl}}\left(
   s \tilde \Omega(s)\ \hat{\vec{e}}_\phi\times{\vec b} - \hat{\vec{e}}_z\times{\vec u}
   \right) + \Delta{\vec b} ,
   \nonumber \\
   {\partial{\vec\omega}\over\partial t} &=&  {\rm Rm\ {\rm curl}}\left(
   s \tilde\Omega(s)\ \hat{\vec{e}}_\phi\times{\vec\omega} -
   {\kappa^2\over 2\tilde\Omega (s)}\ \hat{\vec{e}}_z\times{\vec u}\right)
   \nonumber \\
   &-& {\rm {Pm\ Ha^2\over Rm}\ {\rm curl}}\left(\hat{\vec{e}}_z\times{\vec j}\right)
   + {\rm Pm}\Delta{\vec\omega } ,
\label{8}
\end{eqnarray}
where $\kappa$ is the
normalized epicycle frequency
\begin{equation}
\kappa^2 = {2\tilde\Omega\over s}{{\rm d} s^2 \tilde\Omega\over{\rm d} s} ,
\label{9}
\end{equation}
time and distances are normalized to the diffusive time, $H^2/\eta$, and
the half-thickness, $H$, of the disk.
The boundary conditions on the disk surfaces are (i) stress-free for the
flow and (ii) pseudo-vacuum condition for the magnetic disturbances, i.e.
\begin{equation}
\hat{\vec{e}}_z\times{\vec b} = 0.
\label{5}
\end{equation}
The solutions  are required to be regular at the rotation axis and
to vanish at infinity.

The linear stability analysis with ${\vec b}, {\vec\omega} \sim
{\exp}(\gamma t)$ leads to an eigenvalue problem for the equation
system (\ref{8}). The problem has been solved numerically
with the method developed   by Kitchatinov \& Mazur (1997). As
usual, the system (\ref{8}) allows two types of solutions with
different symmetries about the disk midplane. One of the symmetry
types combines a symmetric magnetic field with an antisymmetric
flow field. This symmetry type is called \lq\lq quadrupolar"  (we
are mainly interested in the symmetry of the magnetic pattern).
The other symmetry type of \lq\lq dipolar" perturbations combines
antisymmetric magnetic fields with symmetric flows. The eigenmodes
are also distinguished by their azimuthal wave number, $m$,
because the axisymmetric modes with $m=0$ and nonaxisymmetric
modes with $m \geq 1$ satisfy independent equation systems.

The primary aim of the linear theory is to define the stability
boundary in parameter space  which separates the
region of stable perturbations with negative or zero real part of $\gamma$
from the instability region with (exponentially) growing
perturbations. The stability boundary depends on the symmetry type of
the excitations. The growth rates of the unstable
disturbances are less significant because they characterize those
states of the system which have little chance to exist just because they are
unstable. The instability region is the realm of nonlinear
simulations.
\section{Results}\label{results}
\begin{figure}
   \resizebox{\hsize}{!}{\includegraphics{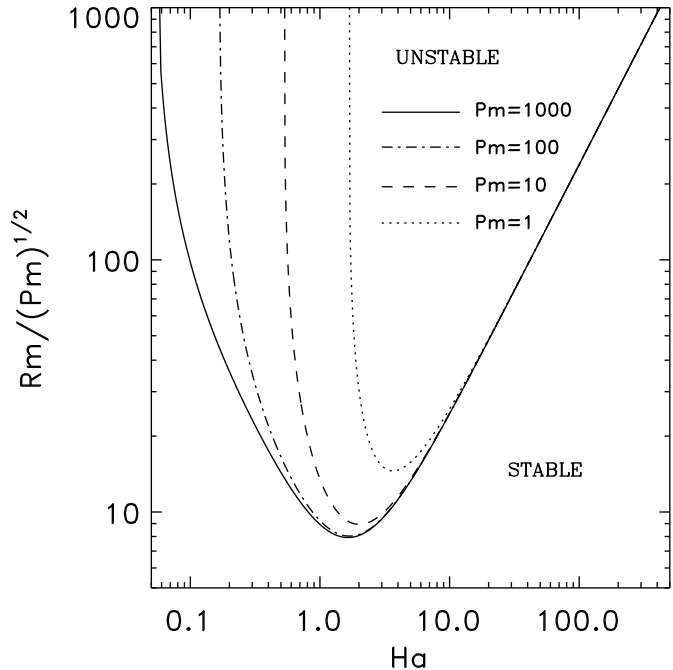}}
   \caption{The stability diagram for axisymmetric modes of quadrupole parity.
            The right branches and minima of the neutral stability lines
            remain constant for  sufficiently large magnetic Prandtl numbers.}
   \label{f1}
\end{figure}

Because of the galactic implications, we focus on the model
results for  large magnetic Prandtl numbers, $\rm Pm\gg
1$. Figure~\ref{f1} shows the neutral stability lines for axisymmetric
quadrupolar modes. As discussed below, modes of this symmetry type are
excited preferentially. The instability region is placed between the left and
right branches of the bounding lines which branches define the minimum and
maximum field strengths for the instability, respectively. The minima of the lines
define the minimum Reynolds number for the instability.

Several neutral stability lines for Pm increasing
from 1 to 1000 are shown in Fig.~\ref{f1}.
Our numerical model cannot match the huge values of Pm expected for galaxies.
However, Eqs.~(\ref{a6})--(\ref{a8}) suggest (and the results of
Fig.~\ref{f1} confirm) that the neutral stability lines approach a
universal behavior in the high Pm limit. The behavior obeys the so-called
similarity rules from which the MRI characteristics for arbitrarily
large Pm can be estimated.

Three similarity rules can be found from the numerical results
illustrated by Fig.~\ref{f1}:

(i) On the right-hand  branch the lines  approach a
constant slope for high Rm, so that one always has
\begin{equation}
   {{\rm Rm}\over{\rm Ha} \sqrt{\rm Pm}}
   \equiv {\Omega_0 H\over V_{\rm A}} \simeq 2.4 .
\label{10}
\end{equation}
Obviously, with $\Omega_0H/2 \simeq c_{\rm s}$ one finds the magnetic Mach
number slightly below unity.

(ii) As Pm grows, the lines approach a common minimum of
\begin{equation}
  \left( {{\rm Rm}\over \sqrt{\rm Pm}}\right)_{\rm min}
  \equiv \left({\Omega_0 H^2\over\sqrt{\nu\eta}}\right)_{\rm min} \simeq 8.
  \label{11}
\end{equation}

(iii) The left-hand  branches on Fig.~\ref{f1} are close to vertical lines
at large Rm. The lines are almost equidistant in the logarithmic scale
suggesting that the Lundquist number S is approaching a constant, i.e.
\begin{equation}
   {\rm S} = \sqrt{\rm Pm}\ {\rm Ha}
   \equiv {V_{\rm A} H\over\eta} \simeq 1.7 .
\label{12}
\end{equation}

The values (\ref{10})--(\ref{12}) provided  by the global model
do not coincide with the estimates (\ref{a6})--(\ref{a8}) from a
local treatment. The difference should be partly attributed to the
different definition of the basic parameters in the local analysis
(cf. eqs.~(\ref{7}) and (\ref{a4})).
However, the scaling used
for large Pm was inferred from the local analysis presented in the
Appendix.

Figure~\ref{f2} shows how the growth rates of unstable
perturbations vary along a horizontal line (constant Rm/$\sqrt{\rm
Pm}$) from the left to the right branches of the stability diagram.
The growth rates are shown in units of the rotation
frequency, $\tau_{\rm rot}^{-1} = \Omega_0/2\pi$, of the disk
core. It is  suggested that as the Reynolds number
increases the lines join together in a common
dependence for different Pm. The typical growth times are of the order
$\tau_{\rm rot}$.
\begin{figure}
   \begin{minipage}[c]{4.35 truecm}
      \includegraphics[width=4.3truecm, height=6truecm]{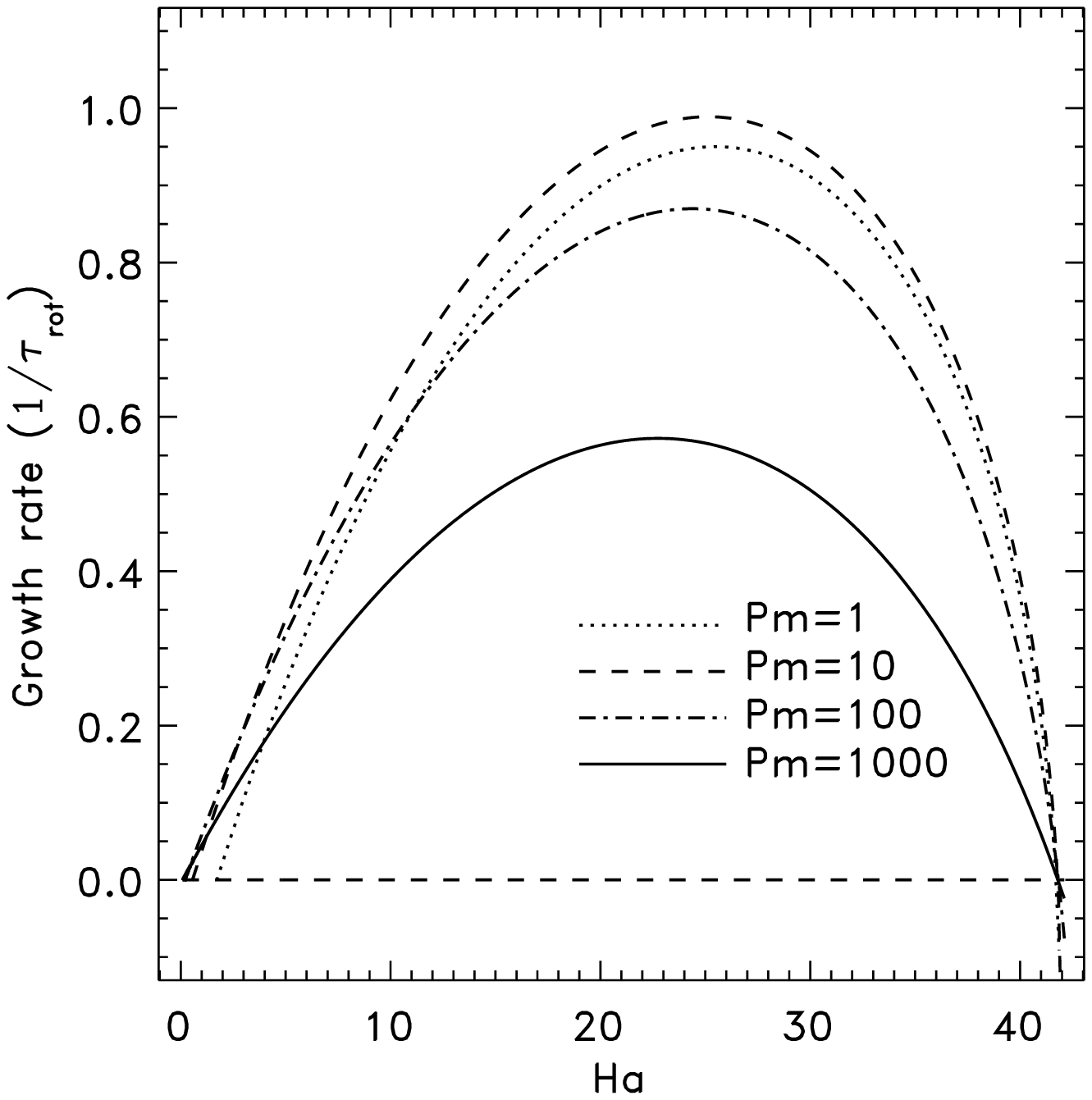}
   \end{minipage}
   \begin{minipage}[c]{4.35 truecm}
      \includegraphics[width=4.3truecm, height=6truecm]{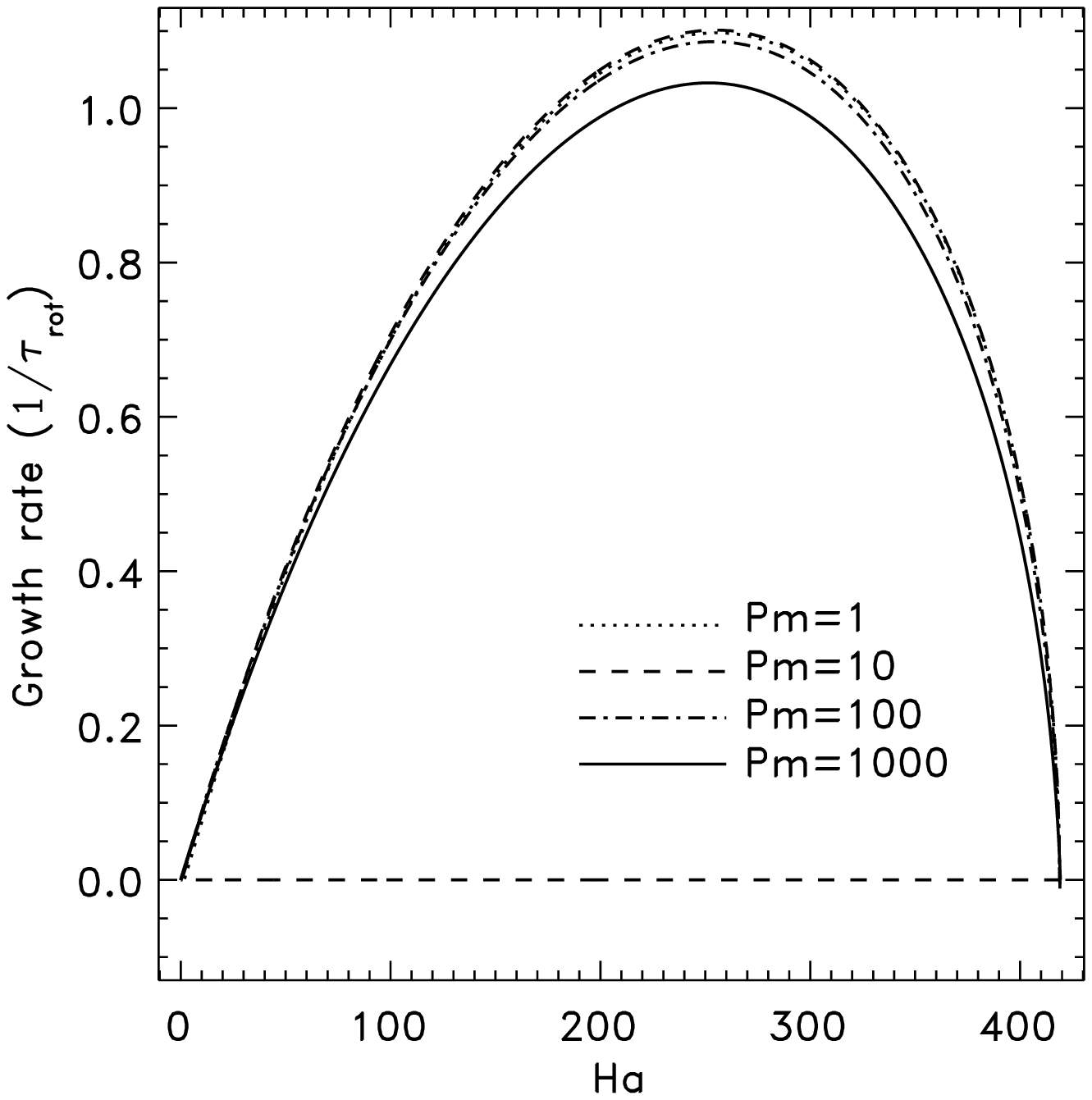}
   \end{minipage}
   \caption{Variation of the normalized growth rates $\gamma \tau_{\rm rot}$
            from the left-hand branch to the right-hand  branch
            across the stability diagram of Fig.~\ref{f1} for
            Rm/$\sqrt{\rm Pm}$ = 100 ({\em left}) and Rm/$\sqrt{\rm Pm}$ = 1000
            ({\em right}).}
   \label{f2}
\end{figure}

\begin{figure*}
\centering
   \begin{minipage}[c]{5.1 truecm}
      \includegraphics[width=5truecm]{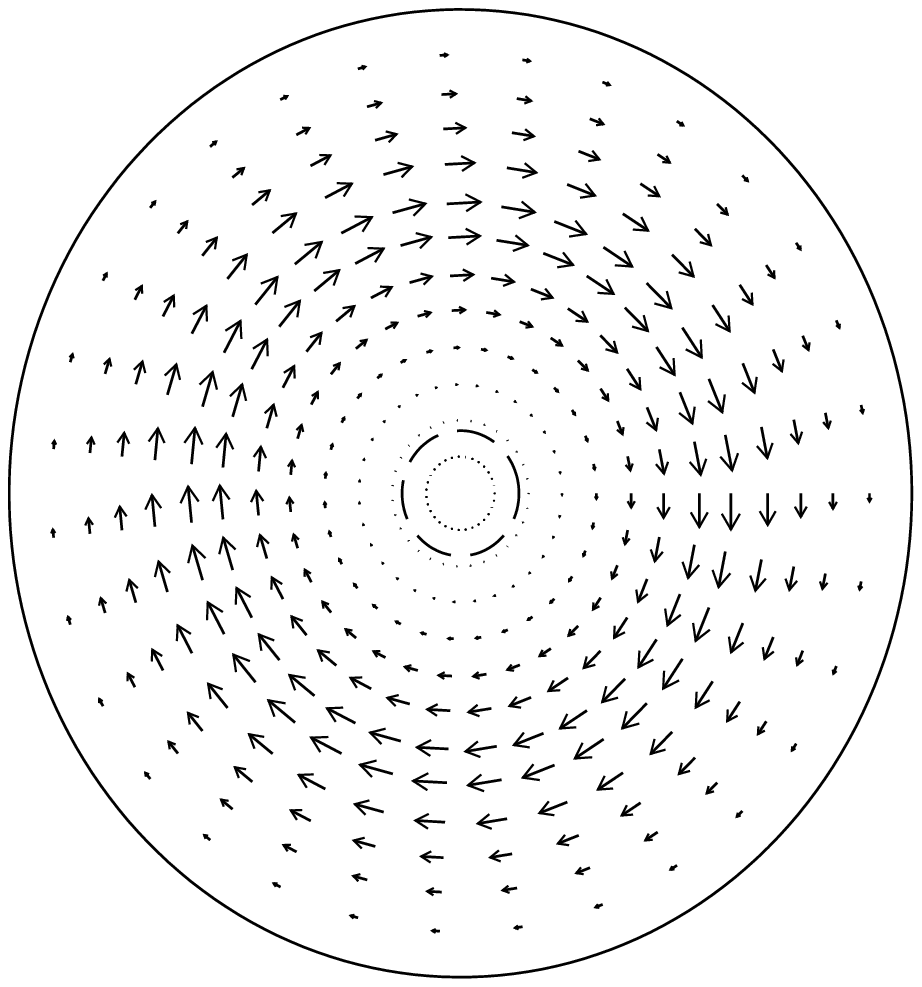}
   \end{minipage}
   \begin{minipage}[c]{5.1 truecm}
      \includegraphics[width=5.truecm]{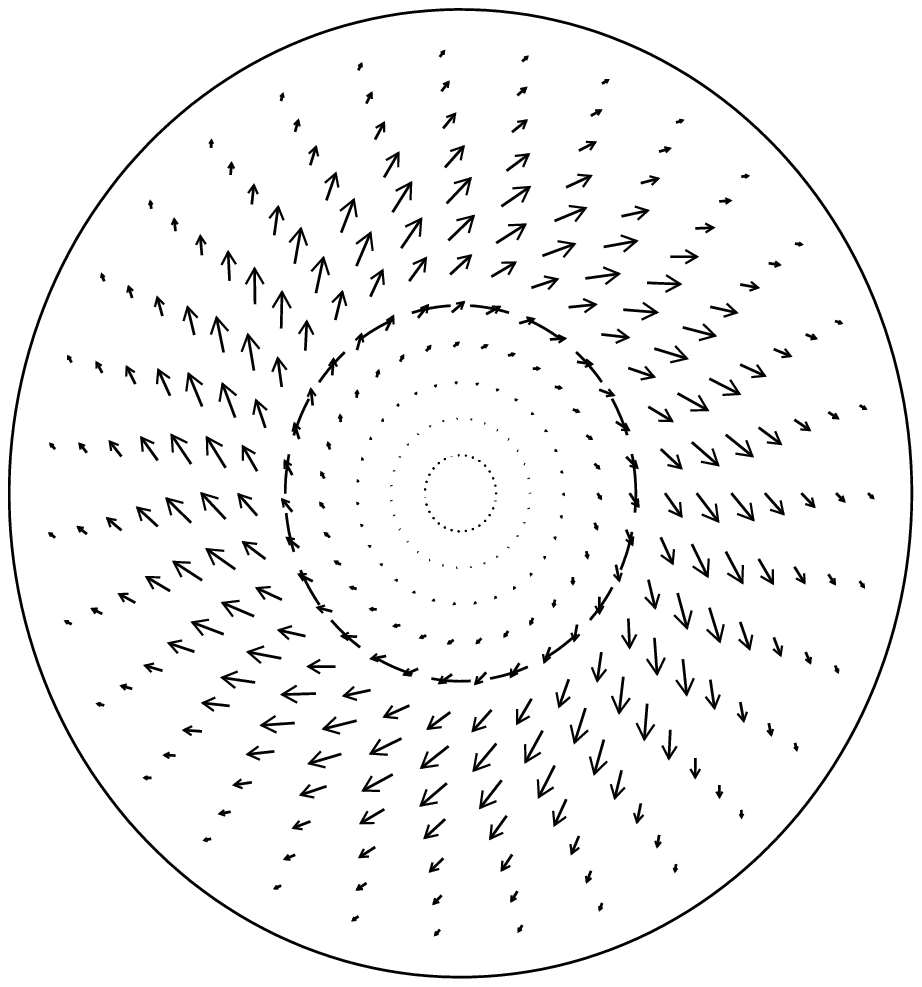}
   \end{minipage}
   \begin{minipage}[c]{5.1 truecm}
      \includegraphics[width=5truecm]{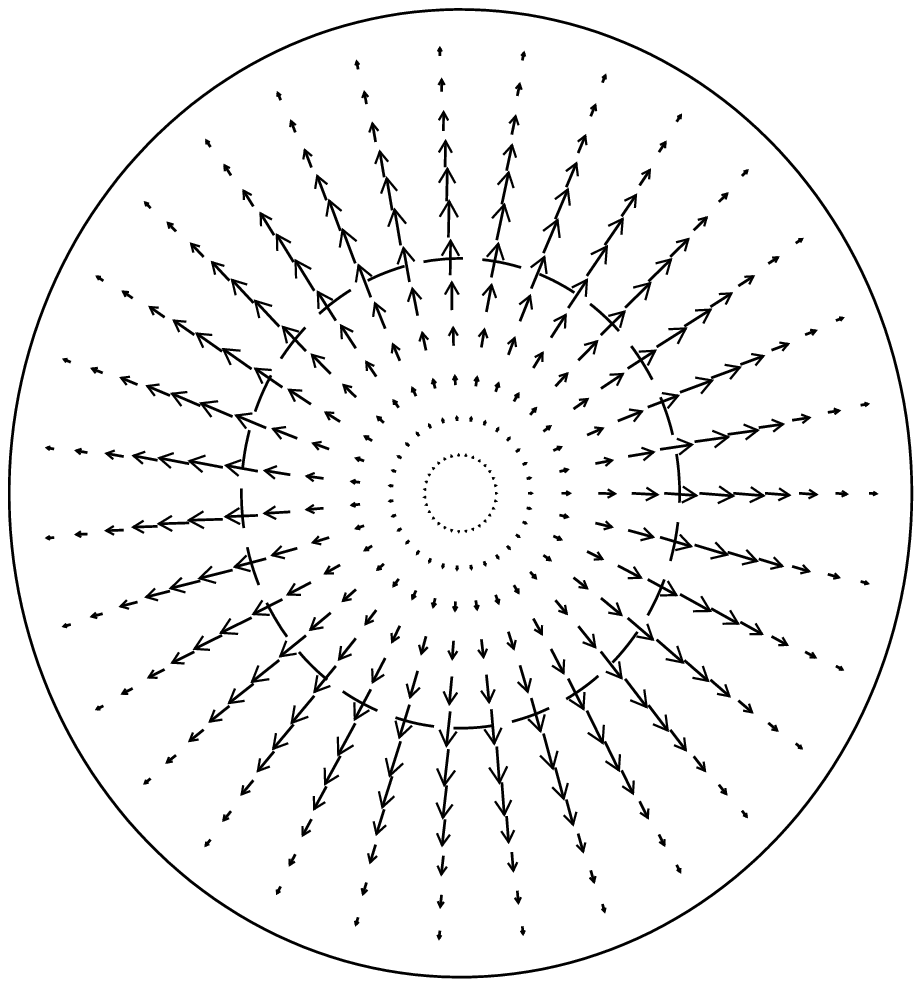}
   \end{minipage}
   \caption{Magnetic field in the midplane of the disk
            for the quadrupole-parity mode at
            Rm/$\sqrt{\rm Pm}=1000$. From the left panel to the right one:
            the marginally stable mode for the minimum magnetic field,
            the mode with maximum growth rate $\gamma = 1.03/\tau_{\rm rot}$
            at ${\rm Ha}\simeq 250$, and the marginally stable mode for maximum
            magnetic field. The broken circle is
            the radius $s_0 = 5H$ of the rigidly rotating core
            of the disk. The background rotation is anticlockwise.
	    Note the increase of the pitch angle and the decrease of the
	    horizontal scale with increasing field strength.}
   \label{f3}
\end{figure*}

Figure~\ref{f3} illustrates the structure of the global modes.
The broken line encircles the rigidly rotating core,
$s_0 = 5 H$. The modes are truly global with large
horizontal scales.
The scale, however, decreases with increasing amplitude of $B_0$.
This behavior is accompanied by an increase of the pitch angle between the
azimuthal direction and the field vector.
The pitch angle is small for the minimum magnetic field.
It increases to almost 90$\degr$ for the maximum
field producing MRI. In both limits the magnetic-induced angular
momentum transport vanishes.

The linear theory cannot yield the magnitude of the magnetic
stress but it can compare it with the kinetic stress. The
angular momentum flux by Reynolds stress is also outward but it is
small compared to the Maxwell stress (cf. Brandenburg et al. 1995;
R\"udiger  et al. 1999). The relation reflects the magnetic nature
of MRI. Also the  energy of the unstable modes is  dominated by
its magnetic part.

\begin{figure}
   \resizebox{\hsize}{!}{\includegraphics{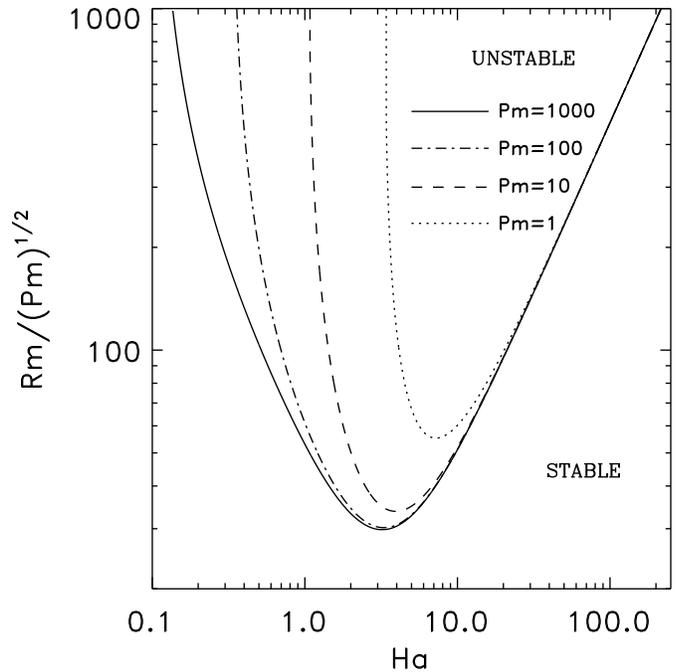}}
   \caption{The stability diagram for dipolar modes. The modes are
   harder to excite than the quadrupolar modes of Fig.~1.}
   \label{f4}
\end{figure}

Figure~\ref{f4} shows the stability diagram for the axisymmetric dipolar
modes. Comparison with Fig.~\ref{f1} reveals a preference
of the quadrupole parity. The neutral stability lines of Fig.~\ref{f4}
lay totally inside the correspondent instability regions of Fig.~\ref{f1}.
This means that when the system approaches instability to the disturbances
of dipolar parity, it is already unstable to quadrupolar perturbations.
In the same sense, the axisymmetric disturbances are preferred compared to
the nonaxisymmetric ones (Kitchatinov \& Mazur 1997).
\section{Implications for galactic dynamos}\label{discussion}
\subsection{Basic parameters}
To probe the relevance of the MRI for galactic magnetism, the parameters
controlling the instability should be estimated for galaxies.

We assume the angular velocity of the disk core as $\Omega_0 \simeq
3\cdot 10^{-15}$~s$^{-1}$ which corresponds to a rotation
period $\tau_{\rm rot} \simeq 70$~Myr (Carignan et al. 1990; Sofue 1996). The
half-thickness of the disk is assumed as $H \simeq 10^{21}\
{\rm cm}\ \simeq\ 300$~pc. For a plasma with not too low degree of ionization,
conductivity is controlled by electron scattering on ions. The magnetic diffusivity
then is,
\begin{equation}
    \eta = 10^{13} T^{-3/2}
    \left({\ln \Lambda\over 20}\right)\ {\rm cm}^2/{\rm s} ,
\label{13}
\end{equation}
where $\ln \Lambda \simeq 20$ is the Coulomb logarithm (Spitzer 1962).
Accepting $T \simeq 10^4$~K for the characteristic temperature, we find
\begin{equation}
   \eta \simeq 10^7\ {\rm cm}^2/{\rm s},\ \ \ \ \ \ \ \ \ \ \ \ \ \ \ \ \ \ \ \ \ \ \
   {\rm Rm}\simeq 3\cdot 10^{20} .
\label{14}
\end{equation}
Viscosity is more difficult to estimate. The expression for the
fully-ionized hydrogen (Spitzer 1962),
\begin{equation}
    \nu = 6.5\cdot 10^7\ {T^{5/2}\over n_{\rm p}}
    \left({20\over\ln \Lambda }\right)\ {\rm cm}^2/{\rm s} ,
\label{15}
\end{equation}
can be applied to H{\sc ii}-regions. With the proton density,
$n_{\rm p} \simeq 1$~cm$^{-3}$, it yields
\begin{equation}
\nu \simeq 7\cdot 10^{17}\ {\rm cm}^2/{\rm s},\ \ \ \ \ \ \ \ \ \ \ \ \ \ \ \ \ \
   {\rm Pm} \simeq 7\cdot 10^{10}.
\label{nu1}
\end{equation}
In the warm interstellar medium, however, neutrals can provide a larger viscosity,
$\nu_{\rm n} \sim v_{\rm t}\ell_{\rm nn}$ ($v_{\rm t}$ is the thermal velocity and
$\ell_{\rm nn}$ is the mean free path for neutral collisions). With the neutrals
density, $n_{\rm n} \simeq 1$~cm$^{-3}$, we find
\begin{equation}
\nu \simeq 10^{21}\ {\rm cm}^2/{\rm s},\ \ \ \ \ \ \ \ \ \ \ \ \ \ \ \ \ \
   {\rm Pm} \simeq 10^{14}.
\label{nu2}
\end{equation}
Fortunately, accuracy in estimating the viscosity is not important. The viscosity
value is needed to ensure only that the minimum ratio (\ref{11}) for the instability
is exceeded. On combining (\ref{14}) with (\ref{nu1}) or (\ref{nu2}), we confirm that
the magnetic Reynolds number exceeds the critical
value resulting from (\ref{11}) by many orders of magnitude. This suggests that
the range of magnetic field strengths producing MRI should be very
broad.

The maximum magnetic field for the instability results from
(\ref{10}),
\begin{equation}
   B_{\rm max} \simeq 6 \ \mu{\rm G} ,
\label{16}
\end{equation}
(cf. Sellwood \& Balbus 1999) which is surprisingly close to the
observed strengths of galactic fields. We applied the estimation of
Alfv\'en velocity, $V_{\rm A} \simeq 2\ B$~km/s with $B$ in $\mu$G,
to obtain (\ref{16}).

The minimum field strength
for MRI can be estimated from (\ref{12}) to the extremely
small value
\begin{equation}
   B_{\rm min} \simeq 10^{-25}\ {\rm G} .
\label{17}
\end{equation}
\subsection{Seed field amplification}
Magnetic field amplification by a hydromagnetic dynamo needs a
seed field to start from. The origin of such a seed field for galaxies is
very uncertain. Three possibilities can currently be found in
the literature.
First, the battery effect of Biermann can produce
the seed field. The corresponding amplitude was estimated by
Davis \& Widrow (2000) and Gnedin et al. (2000) to be about
$10^{-20}$~G in collapsing protogalaxies. Second, collisional friction between
ions and neutrals on ionization fronts may produce fields of
order $10^{-16}$~G (Birk et al. 2002). Third, the cosmological
magnetic field might be as strong as $10^{-10}$~G (Ratra 1992).
All the quoted field strengths can be further amplified by a factor
$\sim 10^2...10^3$ upon the collapsing phase.

Whatever the origin of the seed field might be, the newly formed
galaxies are well inside the instability region between the
limits given in Eqs.~(\ref{16}) and (\ref{17}). The region
should be entered somehow, hence, the instability should start
even before the galaxies are formed. MRI in young galaxies is
operating in far supercritical region, $B\gg B_{\rm min}$.
Growth rates of order $\tau_{\rm rot}$ are expected for this
case. The maximum growth rates initially belong to small
spatial scales, $\ell \ll H$, with $\Omega_0 \ell \simeq V_{\rm
A} \gg \eta/\ell$ (Fricke 1969; Balbus \& Hawley 1991).
The scale range certainly exists
for the very large values given by (\ref{15}). MRI at its nonlinear stage is
known to be able to amplify the magnetic field producing the
instability. As the unstable perturbations
grow, the increase of their magnetic field should shift the
maximum growth rates to larger and larger scales until the
global perturbations start dominating.

The turbulent amplification of magnetic fields was studied by
Beck et al. (1994). They did not specify the origin of turbulence in
favor of quoting observational evidence for the turbulence. We
argue here that the MHD turbulence can be produced by MRI.

If we \emph{assume} that the nonlinear amplification of magnetic
fields proceeds with about the same rate as the
linear growth, the seed field can be amplified by a factor of
$f_{\rm amp}(\tau_{\rm gal}) \simeq \exp\left(\tau_{\rm gal}/\tau_{\rm
rot}\right)$ for a galactic age $\tau_{\rm gal}$. With the value
$\Omega_0 \simeq 3\cdot 10^{-15}$~s$^{-1}$ it takes less than
2~Gyr to produce a 1~$\mu$G field from even the weakest seed
field expected from the Biermann battery effect. This may explain
regular magnetic field of $\mu$G strength
for galaxies with high redshift
$z\simeq 2$ ($\tau_{\rm gal} \sim 1\dots 3$~Gyr). Restrictions on the
magnitude of the seed field, however, would emerge if the microgauss
fields were already present  in very young galaxies  as found at $z \simeq 3.4$
by White et al. (1993).
\subsection{Field geometry}
We have seen in Section~\ref{results}
that MRI produces preferentially the global perturbations of quadrupolar
parity though the background field for the instability is dipolar. Therefore,
MRI supplies the seed field of `correct' parity for global dynamos.
It may be mentioned also that the pitch angles (the
angles between the azimuthal direction and the field vector) in
$\alpha\Omega$-dynamo models are often too small. The small angles
result from an efficient winding of magnetic field by nonuniform rotation.
Figure~\ref{f3} shows that the pitch angles for the most rapidly
growing global modes of MRI are not small.

It is characteristic for linear global MRI models that the symmetries
of flow and field always differ. If the magnetic field is symmetric with
respect to the midplane then the flow is antisymmetric and v.v.
Hence, if really the global MRI is present in galaxies then the
associated flow field is {\em antisymmetric} with respect to the midplane.
Then the effective rotation rate would differ between hemispheres.
Nonlinear calculations must show whether in the final, saturated phase these characteristic mixed-mode properties of MRI-induced magnetism survive.
It might be important in this connection that in the evolving galaxy
with SN-explosions a  second driver of interstellar turbulence occurs which
may finally dominate the  magnetic-field production.
\begin{acknowledgements}
We wish to thank our referee Katia Ferri\`{e}re for constructive comments.
LLK is grateful to AIP for its hospitality and visitor support.
This work was supported by INTAS under grant No.~2001-0550 and by
the Russian Foundation for Basic Research (Project 02-02-16044).
\end{acknowledgements}
\appendix
\section{Local analysis of marginal stability}\label{local}
The guidance for appropriate scaling of our numerical results
came from an elementary local analysis. The local treatment of
MRI provides expectations and better understanding for global
simulations.

We use a Cartesian coordinate system corotating with the local angular
velocity, $\Omega$, and axes $x,y$ and $z$ pointing in the radial, azimuthal
and vertical directions respectively. The local approximation concerns perturbations
with  scales  small compared to the global scale. Then, the  rotation law can be
approximated by the shear flow
${\vec U}_0 = -\hat{\vec e}_y\Omega q x$
where $q$ is the (constant) local shear. The linearized MHD equations
in the local approximation read
\begin{eqnarray}
   {\partial{\vec B'}\over\partial t} &-&B_0{\partial{\vec u'}\over\partial z} -
   x q \Omega{\partial{\vec B'}\over\partial y} + q\Omega B'_x \hat{\vec e}_y
   - \eta\Delta{\vec B'} = 0,
   \nonumber \\
   {\partial{\vec u'}\over\partial t} &+& 2\Omega\hat{\vec e}_z\times{\vec u'}
   - x q \Omega{\partial{\vec u'}\over\partial y} - q\Omega u'_x \hat{\vec e}_y -
   \nonumber \\
   &-& {B_0\over \mu_0\rho}{\partial{\vec B'}\over\partial z} + {1\over\rho}
   \nabla P' - \nu\Delta{\vec u'} = 0 ,
\label{a2}
\end{eqnarray}
(Balbus \& Hawley 1991; Brandenburg et al. 1995) where $P'$ is the pressure
fluctuation including the magnetic term,
$B_0$ is the amplitude of the
background axial magnetic field, ${\vec B}_0 = B_0{\vec e}_z$.

Considering  plane waves with ${\vec B}',{\vec u}',P'\sim {\rm
exp}(\gamma t + {\rm i}k z)$ leads to the dispersion equation,
\begin{eqnarray}
  \lefteqn{\left(\gamma + \eta k^2\right)^2\left( \left(\gamma + \nu k^2\right)^2
   + 2\left(2 - q\right)\Omega^2\right)+}
   \nonumber \\
   &&+ \omega_{\rm A}^2\left( \omega_{\rm A}^2 - 2 q \Omega^2
   + 2\left(\gamma + \nu k^2\right)\left(\gamma +\eta k^2\right)
   \right)
   = 0
\label{a3}
\end{eqnarray}
with the Alfv\'en  frequency $\omega_{\rm A}= k
B_0/\sqrt{\mu_0\rho}$.

We need the boundary separating the regions of
stability and instability in parameter space. The magnetic
Reynolds number and Hartmann number have to be redefined
for the local treatment as
\begin{equation}
   {\rm Rm} = {\Omega\over\eta k^2},\ \ \ \ \ \ \ \ \ \ \ \ \ \ \ \ \  {\rm Ha} = {\omega_{\rm A}\over k^2\sqrt{\eta\nu}} .
\label{a4}
\end{equation}
To find the marginal stability equation we put $\gamma = 0$ in
Eq.~(\ref{a3}). After some algebra it yields
\begin{equation}
   {\rm Rm}^2 = {{\rm Pm}^2\left( 1 + {\rm Ha}^2\right)^2
   \over 2\left( q \ {\rm Pm}\ {\rm Ha}^2 - 2+q \right)} .
\label{a5}
\end{equation}
With a fixed value of Pm, this equation leads to neutral stability
lines similar to those of Fig.~\ref{f1}.
For a given
sufficiently large Rm, the instability exists in the region between some
minimum and maximum values of Ha.
We call the lines defining the
minimum and maximum Ha the  left and the right
branches of the marginal stability line.

Then the  minimum Reynolds number for instability is
\begin{equation}
   {\rm Rm}_{\rm min}^2 = {\rm Pm}{2\over q}\left( 1 +
   {2-q\over q{\rm Pm}} \right)
\label{a66}
\end{equation}
at
\begin{equation}
   {\rm Ha}^2 = 1 + {2\left(2-q\right)\over q {\rm Pm}}.
\label{a6}
\end{equation}
For large Pm, the ratio of Rm/$\sqrt{\rm Pm}$ at minimum approaches
a constant and so does the correspondent Ha.
This behavior at large Pm agrees
with the numerical results of Section~\ref{results}. Actually, the scaling
used in that Section was motivated by the local analysis.

The right branch of the neutral stability lines at high enough Rm
can be found by considering (\ref{a5}) for large Ha,
\begin{equation}
    {\rm Rm}^2 = {{\rm Pm}\over 2 q }{\rm Ha}^2 .
\label{a7}
\end{equation}
Again the ratio of Rm/$\sqrt{\rm Pm}$ is an appropriate
scaling for the slope of the right branch.

The left branch is controlled by the diffusivities. Its expression for large Rm can be obtained by by setting the denominator in (\ref{a5}) to zero.
\begin{equation}
   B_{\rm min}^2  = {2-q\over q}\ \mu_0\rho\eta^2 k^2.
\label{a8}
\end{equation}
The smallest magnetic field producing MRI belongs to the
smallest possible wave number so that also the left
branch is controlled by \emph{global} disturbances.
Equation (\ref{a8}) shows that the minimum Lundquist number is constant on
the left branch for sufficiently high Rm.


\end{document}